\documentclass[twocolumn,showpacs,preprintnumbers,amsmath,amssymb]{revtex4}
\usepackage{graphics}
\begin{document}
%page 1 2nd paragraph hajime no HTSs \to SC cuprates
%2n paragraph end : near \to {\it near}
%the paragraph [In this paper, ....] de, [where T_0 is ...] de [SC] wo huka. 
%page 1 no 3rd paragraph no [16]-->[14,16] to henkoushita (after submission).
%T_g wo define surunowo kyuu-genkou de wasureteita node, VG ... no kasho to 
%t_g no kasho wo wazukani kaeta. 
%Fig.2 no Fig cap. ni symbols no setsumei wo kuwaeta. 
%Eq.(2) mu_n no hyoushiki de L_n no basho no misprint wo correct shita. 

\title{Fluctuation Effects in Underdoped Cuprate Superconductors under Magnetic Field} 

\author{Ryusuke Ikeda}

\affiliation{%
Department of Physics, Kyoto University, Kyoto 606-8502, Japan
}%

\date{\today}

\begin{abstract}
 Fluctuation effects in underdoped cuprates under high fields are examined by trying to fit theoretical results to resistivity and Nernst data in vortex states. The superconducting (SC) fluctuation in underdoped cuprates includes not only the ordinary thermal contribution but also a large amount of quantum dynamical contributions. Together with this, the presence of a SC pseudogap region $T_0-T_{c0}$ increasing with underdoping is found to be the origin of the Nernst coefficient becoming anomalously smaller and the in-plane coherence length {\it apparently} increasing with underdoping. 
\end{abstract}


\maketitle

   It has been well understood that the field-induced fan-shaped 
broadenings of curves of resistivity and thermodynamic quantities, typically seen in optimally (hole-)doped high $T_{c0}$ cuprate superconductors (HTS), are thermal superconducting (SC) fluctuation phenomena mainly in the vortex liquid region of the normal phase {\it below} the zero field ($H=0$) transition point $T_{c0}$ \cite{RI1}. In applied magnetic fields of tesla range perpendicular to the SC planes, the in-plane resistivity and other physical quantities in these materials 
show familiar behaviors \cite{Welp,Huebener} correlated with 
one another. For instance, the onset temperature of fluctuation effects 
suggested from 
resistive data is almost the same as the corresponding one of thermodynamic and thermomagnetic data. This familiar correlation is typically seen in much lower fields than $H_{c2}(0)$, where the fluctuation is purely thermal \cite{Welp}, and the quantum fluctuation contribution is negligible. 

  In contrast, the resistivity in other HTSs with lower $T_{c0}$ often behaves in an uncorrelated manner with thermodynamic quantities. Typically, as the applied field is higher, resistivity data in electron-doped HTSs and some of over (hole-)doped materials show not the fan-shaped \cite{RI1} but a flat curve \cite{Mac1,Naito} following the in-plane normal resistivity $\rho_n(T)=(\sigma_n(T))^{-1}$ curve until a vortex-glass (VG) transition field, lying much below \cite{Naito,Mac2} an effective $H_{c2}(T)$ determined thermodynamically, is approached from above. In over (hole)-doped materials with high $\sigma_n$-value ($\simeq 10^2 (R_q s)^{-1}$ \cite{Mac1}), such an absence of correlation is not surprising because the fluctuation conductivity $\sigma_f$ is negligible compared with $\sigma_n$ in the total conductivity $\sigma=\sigma_n+\sigma_f$ over a wide temperature range, where $R_q = 6.45 ({\rm k \Omega})$ is resistance quantum, and $s \simeq 10({\rm A})$ is a typical size of the spacing between SC layers. However, the corresponding uncorrelated behavior seen in the electron-doped materials \cite{Naito} with $\sigma_n$ of the same order as in the optimally doped YBCO \cite{Huebener} is intrinsic, and its origin needs to be attributed to a fluctuation property. Similar behaviors have been also found in overdoped La$_{2-x}$Sr$_x$CuO$_4$ (LSCO$x$) \cite{Wang} and $\kappa$-(ET)$_2$ organic superconductors \cite{Ito,Sasaki}. As argued elsewhere \cite{RILT23} by fitting to data \cite{Sasaki}, the main origin is expected to consist in the quantum dynamical nature, enhanced with increasing the field, of the SC fluctuation. In general, $\sigma_f$ defined in a Ginzburg-Landau (GL) theory {\it decreases} \cite{RI2} as the SC fluctuation is dominated not by the thermal but rather by a quantum dynamical fluctuation. Such an absence of correlation in the quantum regime {\it near} $T=0$ between the resistivity and the magnetization had been predicted in Ref.11. 

Recent data of resistivity and Nernst coefficient in {\it underdoped} cuprates \cite{Capan,Wang} have also shown similar high field behaviors suggesting a large quantum fluctuation effect. The 2D field-tuned superconductor-insulator transition (FSIT) behavior, seen in resistance data in strongly underdoped cases \cite{Capan,Seidler,Karpinska}, cannot occur without the quantum nature of SC fluctuation \cite{RI2,RI3}, and, as the SC fluctuation is enhanced, the quantum contribution to the fluctuation dominates over the thermal one. Hence, it is natural to expect the SC fluctuation effect to be stronger with underdoping. However, a sharp drop of resistivity in high fields, which often appears even in underdoped materials, was regarded as a mean field -like behavior in the literature \cite{Karpinska,Gant}. Further, the resistance data in the pseudogap regime 
suggest an in-plane coherence length increasing \cite{Ando} with underdoping. 

In this paper, we try to improve understanding on SC fluctuation properties in underdoped cuprates within the GL theory and argue that, together with a quantum fluctuation, a large width $T_0-T_{c0}$ of a SC pseudogap region is a key factor for consistently explaining the conflicting observations in underdoped cuprates, where $T_0$ is the {\it mean field} SC transition temperature 
in $H=0$. 
We start with the 2D GL action expressed in terms of a single component pair-field $\psi({\bf r}, \tau)$  
\begin{eqnarray}
S = \int_{\bf r} \biggl[  \beta \sum_\omega (\psi_\omega({\bf r}))^* \gamma({\bf Q}^2)|\omega| \psi_\omega({\bf r}) \; 
\\ \nonumber
 + \int_0^\beta d\tau \biggl( (\psi({\bf r}, \tau))^* \mu({\bf Q}^2) \psi({\bf r}, \tau) + {b \over 2} |\psi({\bf r}, \tau)|^4 \biggr) \biggr] 
\label{GL}
\end{eqnarray} 
($\hbar$, $k_B=1$), where $\psi(\tau) = \sum_\omega \psi_\omega e^{-i\omega\tau}$, $\beta$ the inverse temperature, $\tau$ the imaginary time, and $b > 0$. The 3D nature due to the coupling between SC planes will be included later in considering a VG fluctuation divergent at $T=T_g$. When the GL approach is applied to the low $T$ and high $H$ region, $H$-dependences of the coefficients $\gamma$, $\mu$, and $b$ need to be taken into account since the familiar low $T$ divergences of these coefficients in $H=0$ clean limit are cut off by the orbital depairing effect of the magnetic 
field (For simplicity, the Pauli paramagnetic depairing effect becoming important in stronger fields will be neglected together with a particle-hole assymmetric dynamical term leading to a fluctuation Hall effect). In fact, the coefficients $\gamma$ and $\mu$ are functionals of the gauge-invariant gradient ${\bf Q}=-{\rm i}\nabla + 2 \pi {\bf A}/\phi_0$ and, once $\psi$ is decomposed into the Landau levels (LLs), are replaced by the coefficients $\gamma_n$ and $\mu_n$ dependent on the LL-index $n$. Hereafter, the familiar clean limit \cite{Lee} will be invoked to describe these coefficients in a form reasonable even in low $T$ and high $H$. Then, $\gamma_n$ and $\mu_n$ are given by 
\begin{eqnarray}
\gamma_n= {\beta \over {2 \pi}} \int_0^\infty ds {s \over {{\rm sinh}(s)}} {\rm L}_n(u_c^2 s^2) \, e^{-(u_c s)^2/2} \;, 
\\ \nonumber
\mu_n={\rm ln}\biggl({T \over {T_0}}\biggr) + \int_0^\infty ds \, {{1- {\rm L}_n(u_c^2 s^2) \, e^{-(u_c s)^2/2}} \over {{\rm sinh}(s)}},
\end{eqnarray}
respectively, where $u_c=T_0 \sqrt{H/(2 H_0 e^\gamma)}/T$, $H_0$ the $T=0$ value of the mean field upper critical field $H_0(T)$ measuring the in-plane coherence length, ${\rm L}_n(x)$ is the $n$-th order Laguerre polynomial, and $\gamma=0.5771$ is the Euler constant. Although, in the $d_{x^2-y^2}$-pairing, cross terms between the lowest LL and the $n=4m$ ($m \geq 1$) higher LLs arise in the quadratic terms of eq.(1), they can be safely neglected in situations of our interest where the lowest LL mode is dominant. The time scales $\gamma_{2m+1}$ vanish at low $T$ limit and are highly sensitive to $T$ and $H$ in contrast to $\gamma_0$ which takes values close to 0.3 in the field and temperature ranges we have examined. Other material and doping dependences will be 
assumed to be included in the coefficient $b$ from which, in low $H$ limit, the $T=0$ magnetic penetration depth $\lambda(0)$ is defined. For instance, (if any) effects of other competing order parameter fluctuations \cite{Fradkin} may be seen as having been integrated out and absorbed into $b$. Further, a numerical computation of $b$ consistent with eqs.(2) suggests that its $H$ and $T$ dependences are similar to those of $\gamma_0$. For these reasons, $b$ will be treated as one of fitting parameters independent of $H$ and $T$. 

To renormalize the $\psi$-fluctuation, the lowest LL approximation will be 
used. Following previous works \cite{RI1,RI2}, the renormalized mass parameter ${\cal G}_0(0)$ defined through the propagator ${\cal G}_0(\omega) = <|\varphi_0(\omega)|^2>=(\gamma_0 |\omega| + ({\cal G}_0(0))^{-1})^{-1}$ for the lowest LL fluctuation field $\varphi_0$ is written as 
\begin{equation}
{\cal G}_0(0) = 1/(\mu_0 + \Delta \Sigma_{\rm h} + \Sigma_0 
+ \Delta \Sigma_{\rm l}). 
\end{equation}
The main roles of $\mu_0$-renormalization are played by the Hartree 
term $\Sigma_0$, which is expressed as  
\begin{eqnarray}
\Sigma_0 = \frac{16 \pi^2 \lambda^2(0)}{\phi_0^2 s \beta} \frac{H}{H_0} \sum_\omega {\cal G}_0(\omega) 
= \frac{16 \pi \lambda^2(0)}{\phi_0^2 s \gamma_0} \frac{H}{H_0} \\ \nonumber 
\times \int_0^{\epsilon_c} d\epsilon \ {\rm coth}\biggl(\frac{\beta \epsilon}{2 \gamma_0} \biggr) \ 
\frac{\epsilon}{\epsilon^2+({\cal G}_0(0))^{-2}} 
\label{Sigma}, 
\end{eqnarray}
where the cutoff $\epsilon_c$ is a constant of order unity, and the coefficient $b$ was replaced with the familiar GL expression \cite{RI1} $16 \pi^2 \lambda(0)^2/(\phi_0 H_0)$. Although other renormalization (correction) term $\Delta \Sigma_l$ within the lowest LL (see, for instance, eq.(2.11) of Ref.11) should be included, it is not essential to our semiquantitative comparison with the data and may be dropped hereafter. The term $\Delta \Sigma_h$ expressing a sum of higher LL contributions is insensitive to $H$ at least in $H \ll H_0$ and can be regarded as contributing to a shift of $H=0$ transition temperature in $\mu_0$. Then, $\Delta \Sigma_h$ may be written 
as ${\rm ln}(T_0/T_{c0})$ \cite{RI4}, and the {\it effective} upper critical field $H_{c2}^*(T)$, defined consistently with $T_{c0}$, is determined in the clean limit by $\mu_0 + \Delta \Sigma_h =0$ and takes the form 
\begin{equation}
H_{c2}^*(T) = H_0 \biggl({{T_{c0}} \over {T_0}} \biggr)^2 \Phi(t),
\end{equation}
while $H_0(T)=H_0 \Phi(T/T_0)$, where $t=T/T_{c0}$, and the 
function $\Phi(x)$ satisfies $\Phi(0)=1$ and $\Phi(1)=0$. 
If $1-T_{c0}/T_0 \ll 1$, as in optimally doped YBCO \cite{RI1}, the presence of the parameter $(T_{c0}/T_0)^2$ is unimportant in eq.(5), and physical properties below $T_{c0}$ may be described without distinguishing $T_{c0}$ from $T_0$. However, in cases with a large $T_0/T_{c0}$, this parameter significantly affects fluctuation phenomena in nonzero fields. 

Now, let us consider transport quantities. Although, in addition to the lowest LL mode, the $n=1$ LL mode and the in-plane electric current (EC) vertices need to be examined to obtain $\sigma_f$ and the transport entropy $s_\phi$, microscopic consideration on $\sigma_f$ can be avoided as follows. As shown previously \cite{RI3}, the mean field vortex flow property requires that, irrespective of microscopic details, the $n=1$ {\it renormalized} mass parameter ${\cal G}_1(0)$ below $H_{c2}^*(T)$ should be given by a factor accompanying the EC vertex. On the other hand, since ${\cal G}_1(0)$ below $H_{c2}^*(T)$ is well approximated by $(\mu_1-\mu_0)^{-1}$ \cite{RI4} insensitive to $T$ at low $T$, the EC vertex is found without microscopic calculations. 
Then, $\sigma_f$ calculated in terms of the Kubo formula consistently with eq.(3) is \cite{RI2} 
\begin{eqnarray}
s R_q \sigma_f = {\gamma_0 \over {2 ({\cal G}_1(0))^2 \beta}} \sum_\omega \biggl[ {\cal G}_0(\omega) {\cal G}_1(\omega) ( {\cal G}_0(\omega) 
\\ \nonumber
+ g {\cal G}_1(\omega) ) - {{({\cal G}_0(\omega))^2 + g^2 ({\cal G}_1(\omega))^2} \over {({\cal G}_1(0))^{-1} + g ({\cal G}_0(0))^{-1}}} \biggr],
\end{eqnarray}
where $g=\gamma_1/\gamma_0$, and ${\cal G}_1(\omega)=(\gamma_1|\omega|+({\cal G}_1(0))^{-1})^{-1}$. It is easily seen that, in the quantum ($T \to 0$) limit, eq.(6) vanishes \cite{RI2} and that, in the opposite thermal limit with no $\omega \neq 0$ terms, eq.(6) is independent of ${\cal G}_1(0)$ due to the relation $g {\cal G}_1(0) \ll {\cal G}_0(0)$. Further, we numerically verified that, even if the $\omega \neq 0$ terms are included, this cancellation on ${\cal G}_1(0)$ works 
extremely well particularly in higher $H$. 

On the other hand, $s_\phi$ is, by definition \cite{Dorsey}, proportional to the heat current vertex, which may have a strong $T/T_0$ dependence of electronic origins. For brevity, we use hereafter the GL expression \cite{Dorsey} of the heat current and will not consider the very low $T/T_0$ region in which $s_\phi$ decreases \cite{Maki} upon cooling independently 
of $\sigma$ (see the figures). Consistently with eq.(6), $s_\phi$ is obtained in terms of a Kubo formula and, using $g {\cal G}_1(0) \ll {\cal G}_0(0)$, is simplified as  
\begin{equation}
s_\phi \simeq \frac{H}{H_0 s {\cal G}_1(0)} \sum_\omega {\cal G}_1(\omega) {\cal G}_0(\omega) \simeq \frac{\phi_0^2}{16 \pi^2 \lambda^2(0) T} \Sigma_0,  
\end{equation}
where the factor $({\cal G}_1(0))^{-1}$ is carried by an ET vertex. Namely, in the lowest LL approximation, $s_\phi$ in the GL region is proportional to the fluctuation entropy even in the quantum case, and the mean field 
result $\phi_0^2 \beta (1-T/T_{c0})/(16 \pi^2 \lambda^2(0))$ is recovered when both $H$ and $T$ are lowered enough . 

%In fact, we have found by deliberately 
%changing $\mu_{1{\rm R}}$-value that making the $mu_{1{\rm R}}$-value half 
%leads to, more or less, a couple of percents of changes in the computated 
%conductance. Therefore, we will express $\mu_{1{\rm R}}$ and the factor of the %current vertex as $\mu_1-\mu_0$. 
We have tried to fit theoretical curves following from eqs.(6) and (7) to the resistivity $\rho$ and Nernst signal data in LSCOx samples with $x=0.06$ \cite{Capan2} and $0.08$ \cite{Capan}, and the results are given in Figs.1 and 2, respectively, where the Nernst coefficient $N=\rho s_\phi/\phi_0$. The used values of material parameters are shown in the figure captions. Since, by definition, our $T_{c0}$ in 2D corresponds to a (amplitude-) fluctuation-corrected BCS critical temperature denoted in Ref.\cite{HN} as $T_c^0$, $T_{c0}$ was identified in the figures with the onset of a remarkable resistivity drop in $H=0$. 
The normal conductivity $\sigma_n$ is assumed to take the empirical form ${\rm const}./{\rm ln}(T_p/T)$ \cite{Boeb} with $T_p \gg T_{c0}$. 
Regarding the VG fluctuation term $\sigma_{vg}$, added to $\sigma$ 
for describing the low $T$ tails of $\rho$-curves, a 3D form $\sigma_{vg}= 0.01 (R_q s)^{-1} \gamma_0 T_{c0}/(t-t_g)^4$ was assumed in Fig.2 with $t_g \equiv T_g/T_{c0}=0.125$ ($0.016$) for 
12 (26) (T), while a 2D form \cite{RI3} used in analyzing the FSIT behaviors in the $s$-wave pairing case was applied in Fig.1 by assuming a vortex pinning 
strength as a fitting parameter independent of $\lambda(0)$. Details of such an analysis of FSIT behaviors will be explained elsewhere \cite{RILT23}. We simply note here that the detailed forms of $\sigma_{vg}$ are inessential to our main conclusion given below. For instance, the flat $\rho$-curves near 4(T) in Fig.1 are created mainly by a quantum behavior \cite{RI2} eq.(6) shows, and $\sigma_{vg}$ contribution was quantitatively 
negligible there. 

Although there is a slight disagreement in $T$-dependences of $N$ between the data and theoretical curves, we feel that the fitting to the data is satisfactory when taking account of the use of our simplified model with a reduced number of fitting parameters. The figures show an enhancement of {\it quantum} SC fluctuation accompanying the underdoping. In Fig.2, $H_{c2}^*(0)$ is close to 26 (T), while $H_{c2}^*(4 ({\rm K})) \simeq 8$ (T) and  $H_{c2}^*(7({\rm K})) \simeq 6$ (T) in Fig.1 so that the resistance may show an insulating behavior even below $H_{c2}^*(T)$. Further, the fitting results imply the following doping dependences of material parameters. First, $\lambda(0)$ significantly increases with 
underdoping and, in $x=0.06$ case, is longer than 1($\mu$m). Judging from the LSCO data \cite{Pana}, the $\lambda(0)$-values used in the figures are not 
unreasonable. Secondly, through the doping dependence of $H_0$, the in-plane coherence length {\it decreases} with underdoping in contrast to the experimental estimation \cite{Ando}. Further, the SC pseudogap region measured by $T_0-T_{c0}$ was assumed to become wider with underdoping. It is unclear whether the obtained $T_0$-value should be {\it quantitatively} compared with, for instance, the onset temperature $T_\nu$ \cite{Wang2} of Nernst effect. Actually, the $T_0=96$(K) in Fig.2 was estimated from the data for larger $H/H_0$ values than in Fig.1, and its actual value may be slightly higher. Thus, a doping dependence of $T_0$ suggested through the figures does not necessarily contradict $T_\nu$ in LSCO \cite{Wang2} {\it decreasing} with underdoping in $x \leq 0.1$. 

It is important to notice that the $\sigma_f$-expression of eq.(6) is invariant under the replacement of parameters, $\lambda(0) \to \lambda(0) T_{c0}/T_0$, $H_0 \to H_{c2}^*(0)$, and $T_0/T_{c0} \to 1$. Namely, the presence of a large SC pseudogap 
is not uncovered by examining {\it only} magnetoresistance data, and a 
neglect of SC pseudogap region would lead to a much shorter penetration depth and an in-plane coherence length {\it growing} \cite{Ando} with underdoping. More importantly, as a result of the much shorter penetration depths, the assumption $T_0 = T_{c0}$ leads to $N$-values in $x=0.08$ case which are one order of magnitude larger than the data in Fig.2 and to $N$-values in $x=0.06$ case which are two order of magnitude larger than in Fig.1 !  It is quite difficult to resolve such a serious discrepancy, for instance, simply by improving the prefactor of the heat current. This result, requiring a $T_0-T_{c0}$ increasing with underdoping, agrees with the opinion that $T_\nu$, much higher than $T_{c0}$, is essentially identical with the {\it mean field} SC transition point $T_0$ or $H_0(T)$. 

Recently, Wang et al. \cite{Wang} 
have argued that a field $H^*$ at which Nernst 
coefficient reaches its maximum arises, in their overdoped LSCO, from a simultaneous sharp drop of $\rho$ related to an effectively longer vortex core size. From Fig.1, such a {\it mean field} argument on the sharp drop of $\rho$ is generally invalid, since a picture based on an anomalous feature near the vortex cores would become more applicable with underdoping, while a comparison between the $\rho$ and $N$-data in high fields in Fig.1 clearly shows that the decrease of $N$ at lower temperatures is due not to $\rho$ but to $s_\phi$. As mentioned in Introduction, a sharp drop of $\rho$ much below $H_{c2}^*(T)$ ($\leq H_0(T)$) occurs even in organic materials \cite{Sasaki,RILT23} and is due not to a mechanism peculiar to the 
cuprates but to a 3D VG transition in systems with a large 
quantum SC fluctuation. 

In conclusion, the resistance and Nernst data in underdoped LSCO were consistently explained to clarify the doping dependences 
of fluctuation effect and of SC parameters. The in-plane coherence length was argued to decrease with underdoping. To explain those transport data consistently, microscopic details near the vortex cores are not necessary, and taking account of the quantum SC fluctuation and a SC pseudogap region $T_0-T_{c0}$, both of which are enhanced with underdoping, is indispensable. 

The author thanks C. Capan and W. Lang for sending him their unpublished data 
and for useful discussions. 

%\bibliographystyle{unsrt}

%\vfil\eject
\begin{figure*}[ht]
\begin{center}\leavevmode
\begin{minipage}[t]{8cm}
\includegraphics{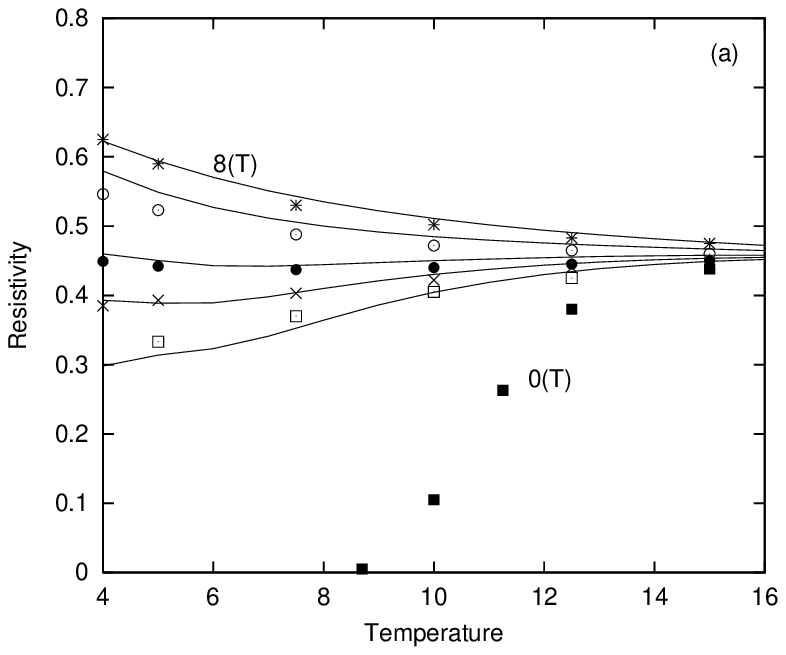}
\includegraphics{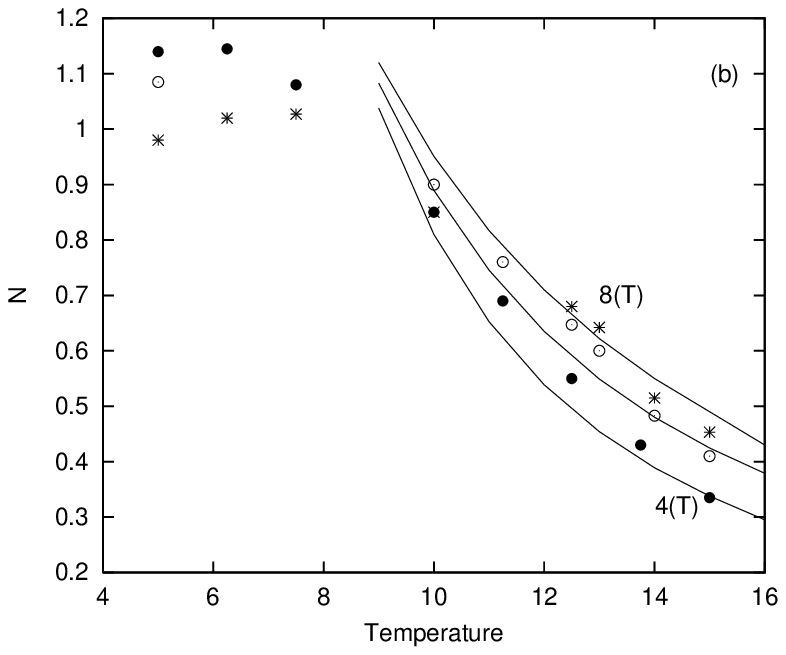}
\end{minipage}
\caption{\label{Fig.1} (a) Resistivity $\rho$ (m${\rm \Omega} \cdot$cm) and (b) Nernst coefficient $N$ ($\mu$V/K) data in LSCO x=0.06 \cite{Capan2} in 2 (open square), 3 (cross), 4 (closed circle), 6 (open circle), and 8 (asterisk)(T) at each $T$ (K) and the corresponding theoretical (solid) curves. The used parameter values are $\lambda(0)=2.3$($\mu$m), $H_0=493$(T), $T_{c0}=13$(K), $s=1.5$ (nm), and $T_0=96$(K).}
\end{center}\end{figure*}

\begin{figure*}[ht]
\begin{center}\leavevmode
\begin{minipage}[t]{8cm}
\includegraphics{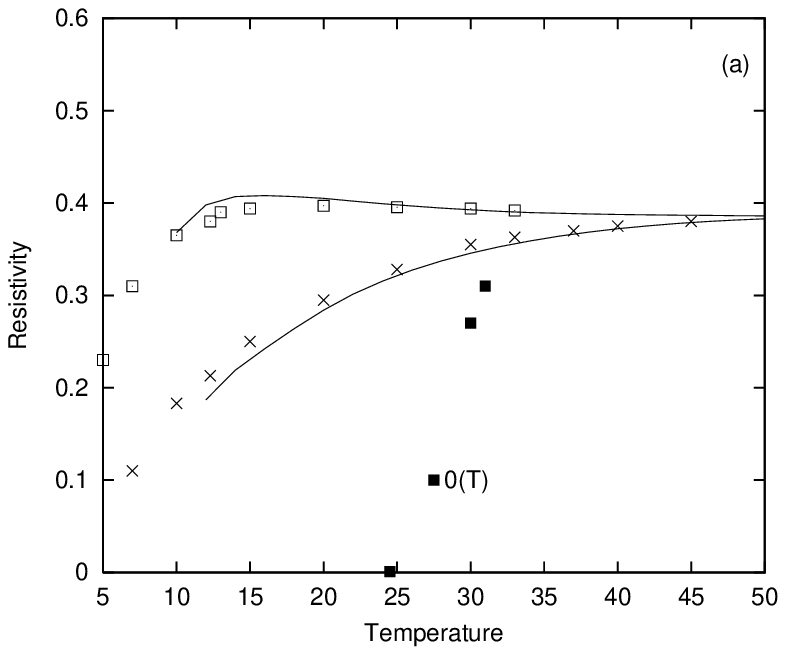}
\includegraphics{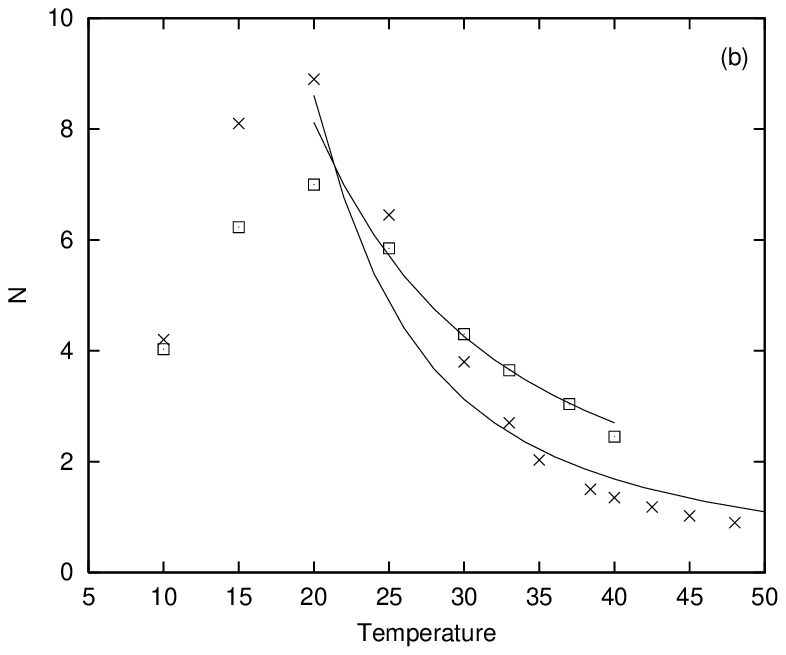}
\end{minipage}
\caption{\label{Fig.2} Corresponding results to Fig.1 for LSCO x=0.08 \cite{Capan} in 12 (cross) and 26 (open square) (T). The parameter values are $\lambda(0)=0.43$($\mu$m), $H_0=235$(T), $T_{c0}=32$(K), $s=1.5$ (nm), and $T_0=96$(K).}
\end{center}\end{figure*}


\begin{thebibliography}{99}
\bibitem{RI1} R. Ikeda et al., J. Phys. Soc. Jpn. {\bf 60}, 1051 (1991); Phys. Rev. Lett. {\bf 67}, 3874 (1991). 
\bibitem{Welp} U. Welp et al., Phys. Rev. Lett. {\bf 67}, 3180 (1991). 
\bibitem{Huebener} H.-C. Ri et al., Phys. Rev. B {\bf 50}, 3312 (1994). 
\bibitem{Mac1} A. P. Mackenzie et al., Phys. Rev. Lett. {\bf 71}, 1238 (1993). 
\bibitem{Naito} S. Kleefisch et al., Phys. Rev. B 63, 100507 (2001) and references therein. 
\bibitem{Mac2} A. Carrington et al., Phys. Rev. B {\bf 54}, R3788 (1996). 
%\bibitem{Wen} H.H. Wen et al., Phys. Rev. Lett. {\bf 82}, 410 (1999). 
%\bibitem{Gesh} V. G. Geshkenbein, L. B. Ioffe and A. J. Millis, Phys. Rev. 
%Lett. {\bf 80}, 5778 (1998). 
\bibitem{Wang} Y. Wang et al., Phys. Rev. Lett. {\bf 88}, 257003 (2002).
\bibitem{Ito} H. Ito et al., Physica B {\bf 201}, 470 (1994).  
\bibitem{Sasaki} T. Sasaki, private communication; M. Lang et al., 
Phys. Rev. B {\bf 49}, 15227 (1994). 
\bibitem{RILT23} R. Ikeda, presented in LT23 (to appear in Physica C) and 
in preparation. 
\bibitem{RI2} R. Ikeda, Int. J. Mod. Phys. B {\bf 10}, 601 (1996). 
\bibitem{Capan} C. Capan et al., Phys. Rev. Lett. {\bf 88}, 056601 (2002). 
\bibitem{Seidler} G. T. Seidler et al., Phys. Rev. B {\bf 45}, 10162 (1992). 
\bibitem{Karpinska} K. Karpinska et al., Phys. Rev. 
Lett. {\bf 77}, 3033 (1996). 
\bibitem{RI3} H. Ishida and R. Ikeda, J. Phys. Soc. Jpn. {\bf 71}, 254 (2002). 
\bibitem{Gant} V. F. Gantmakher et al., JETP {\bf 88}, 148 (1999). 
\bibitem{Ando} Y. Ando and K. Segawa, Phys. Rev. Lett. {\bf 88}, 167005 
(2002). 
\bibitem{Lee} See, for instance, Appendix in P. A. Lee and M. G. Payne, 
Phys. Rev. B {\bf 5}, 923 (1972). 
%\bibitem{Dorsey} S. Ullah and A. T. Dorsey, Phys. Rev. B {\bf 44}, 
%262 (1991). 
\bibitem{Fradkin} S. A. Kivelson et al., cond-mat/0205228. 
\bibitem{RI4} R. Ikeda, J. Phys. Soc. Jpn. {\bf 64}, 1683 (1995). 
%\bibitem{RI5} R. Ikeda, J. Phys. Soc. Jpn. {\bf 70}, 219 (2001). 
\bibitem{Dorsey} R. J. Troy and A. T. Dorsey, Phys. Rev. B {\bf 47}, 2715 
(1993). 
\bibitem{Maki} On this issue in the $s$-wave pairing case, see 
K. Maki, Physica (Utrecht) {\bf 55}, 125 (1971). 
\bibitem{Capan2} C. Capan, private communication.  Regarding $x=0.06$ case, we have prefered the refined data shown in Fig.1 to those in Ref.12 with a very broad resistive transition in $H=0$. 
\bibitem{HN} B. I. Halperin and D. R. Nelson, J. Low Temp. Phys. {\bf 36}, 599 (1979). 
\bibitem{Boeb} S. Ono et al., Phys. Rev. Lett. {\bf 85}, 638 (2000). 
\bibitem{Pana} C. Panagopoulos et al., cond-mat/0007158. See Fig.3 there. 
\bibitem{Wang2} Y. Wang et al., Phys. Rev. B {\bf 64}, 224519 (2001).  
%\bibitem{Matsuda} T. Nagaoka et al., Phys. Rev. Lett. {\bf 80}, 3594 (1998). 
%\bibitem{RI6} R. Ikeda, Physica C 316, 189 (1999). As indicated there, the Hall%-sign reversal near $T_{c0}$ in underdoped materials is rather natural if the d%oping dependence of $T^*$ is opposite to that of $T_{c0}$. 

\end{thebibliography}
\end{document}